\documentclass[aps,showpacs,a4paper,floatfix,twocolumn,prc,amsmath,amssymb]{revtex4-1}
\usepackage{graphicx}
\usepackage{epsfig,hyperref}
\usepackage{ulem,color}
\usepackage{morefloats}
\usepackage{rotating}
\usepackage{relsize}
\usepackage{url}

\begin{document}

\title{ Crust-core transition of a neutron star: effects of the symmetry energy and
temperature under strong magnetic fields}

\author{Jianjun Fang}
\author{Helena Pais}
\author{Sagar Pratapsi}
\author{Constan\c ca Provid{\^e}ncia}

\affiliation{CFisUC, Department of Physics, University of Coimbra,
  3004-516 Coimbra, Portugal.}

\begin{abstract}
We study the simultaneous effects of the symmetry energy and temperature on the crust-core
transition of a magnetar. The
dynamical and the thermodynamical spinodals are used to calculate the
transition region within a relativistic mean-field approach for the
equation of state.
Quantizing magnetic fields with intensities in the range of  $ 2\times 10^{15}<B<5\times 10 ^{16}$G
are considered. Under these 
strong magnetic fields, the crust extension is very
sensitive to the density dependence of the symmetry energy, and the
properties that depend on the crust thickness could set a constraint on
the equation of state. It is shown that the
effect on the extension of the crust-core transition is washed out for
temperatures above $10^{9}$ K. However, for  temperatures below that value, a noticeable
effect exists that grows as the temperature decreases and which  should be taken into account when the  evolution
of magnetars is studied.
\end{abstract}

\pacs{24.10.Jv,26.60.Gj,26.60.-c}
\maketitle

{\it Introduction:} Magnetars are isolated neutron stars identified as
x-ray pulsating sources and soft $\gamma$-ray repeaters with very strong surface magnetic fields,
$B=10^{14}-10^{15}$ G, and long spin periods ($P=1-12$ s). Presently,  almost thirty magnetars have been identified, see
\cite{kaspi14,sgr}.

The long term evolution of magnetars has been carried out  in
Ref. \cite{pons13}. The authors found that for high values of  a temperature independent  impurity parameter  considered in the upper layers of the inner crust, where
according to \cite{pethick83} pasta phases could occur,
an enhanced dissipation of the magnetic field
is maintained, causing a fast spin down rate of the star. This could be
the reason for the non-detection of  isolated neutron stars with
periods above 12 s.  How properties of the pasta phase affect  electrical and
thermal conductivities is still not clear \cite{horowitz15,yakovlev15}.

The crust equation of state (EoS) and its extension, together with the crust-core transition region  
 seem also to play a central role in the 
evolution of the magnetar magnetic field, in the determination of its configuration
\cite{lander14,lander16}, and in the description of the observed
quasiperiodic oscillations 
(QPO) of soft $\gamma$-ray repeaters \cite{gabler13}. Also other crust properties, such as the neutron-drip
 transition that characterizes the outer-inner crust transition, or
 the outer-crust structure and composition  are
 affected by strong magnetic fields \cite{chamel15,basilico15}.

In  \cite{fang16,fang17}, the effect of
strong magnetic  fields on the inner crust of neutron stars was
discussed within a relativistic mean-field (RMF) model, and  several
interesting results were obtained. It was found  that
the inner crust is more complex in the presence of strong magnetic
fields, and alternating regions of clusterized and non-clusterized
matter appear above the $B=0$ crust-core transition density. Contrary
to the $B=0$ case, the crust-core transition is defined by a  region with a nonzero density
width for magnetic fields above $ \sim 10^{15}$ G. It
was also shown that the width of the transition region is sensitive to the model.
This transition region could support the possible existence of  highly resistive matter 
at the upper layers of the inner crust that
enhances the decay of the magnetic field. 

 Neutron star glitches is another phenomenon 
explained by the crust properties 
\cite{link99}. The crust fractional momentum of inertia is a  crucial
quantity to interpret glitches. However,  recent works have pointed
out that due to entrainment effects, that couple the
superfluid neutrons to the solid crust, the crust would not be enough
to describe glitches \cite{chamel13,glitch2}. The increase of the
inner crust due to magnetic field effects found in Refs. \cite{fang16,fang17} could validate the crustal
contribution to the  description of the  glitch mechanism.

The cooling of the inner crust of a neutron star occurs more
slowly  than the core, where a direct Urca process may
originate a very fast cooling. During the first years of the star, the
cooling of the outer crust, inner crust and core occur
independently. It is only when the star is $\sim 50$yr old that its
total relaxation has occurred \cite{yakovlev01}. The
temperature of the crust depends on the star mass and on the EoS, but a newly
born star,  less than one year old, will have a temperature 
above 10$^9$ K.  At the star's total relaxation, the temperature has
dropped well below  $\sim 10^9$ K. Moreover, 
the magnetic field and temperature evolutions are strongly coupled in a
neutron star which require coupled magneto-thermal evolution to
properly study the star cooling \cite{aguilera08,vigano12,vigano13}. It is,
therefore, of interest  to study how sensitive is the
increase of the crust-core transition region  to 
temperature.

In the present study, we will use RMF models \cite{walecka}, which are phenomenological models constrained by different types of observables, in particular,
experimental measurements, theoretical ab-initio calculations and
observations in astronomy, see \cite{oertel17} for a review. Taking a set of models that have the same isoscalar
properties at saturation, and only differ on the isovector properties,
will allow to investigate how the effect of the magnetic field on  
the stellar matter  depends
on the properties of the EoS,  in particular, on the
density dependence of the symmetry energy.

First we  analyze  the  effect of the
 density dependence of the
 symmetry energy on the  magnetar crust-core transition within the dynamical
 spinodal formalism at zero temperature. Next, the effect of temperature  
is studied. This will be done using  the  finite temperature
 thermodynamical spinodal \cite{chomaz03,avancini06}, and temperatures between 1 and 1000 keV
 ($10^7-10^{10}$ K). 
 Although the crust-core
 transition density is $\sim 10\%$ larger in the thermodynamical
 spinodal approach, as compared to the dynamical one, see
 \cite{ducoin10,ducoin11}, we believe it will allow us to perform a realistic
 discussion. The same approach was used at zero temperature to study
 the liquid-gas phase transition of magnetized nuclear matter in \cite{aziz08,chen17}.

{\it Formalism:} Stellar matter is described within the nuclear RMF formalism under the effect of strong
magnetic fields \cite{broderick,aziz08}. The anomalous magnetic moment
(AMM) is included in part of the calculations. The nuclear interaction
is described through the inclusion of mesonic fields: an isoscalar-scalar field $\phi$ with mass $m_s$, an isoscalar-vector
field $V^{\mu}$ with mass $m_v$, and an isovector-vector field
$\mathbf b^{\mu}$ with mass $m_\rho$. Besides nucleons with mass $m$,
electrons with mass $m_e$ are also included in the Lagrangian density. Protons and electrons interact through the electromagnetic field
$A^{\mu}$, which  includes a static component  assumed to be
externally generated, $A^{\mu}=(0,0,Bx,0)$, 
so that $\mathbf{B}=B\, \hat{z}$ and $\nabla \cdot {\bf A}$=0. 
We take the usual  RMF   Lagrangian density  
$
{\cal L}=\sum_{i=p,n} {\cal L}_i + {\cal L}_e + \cal L_\sigma + {\cal
  L}_\omega + {\cal L}_\rho + \mathcal{L}_{\omega \rho }+ {\cal L}_{A},$
where ${\cal L}_i$ is the nucleon Lagrangian density, given by
$$
{\cal L}_i=\bar \psi_i\left[\gamma_\mu i D^\mu-M^*_i-\frac{1}{2}\mu_N\kappa_b\sigma_{\mu \nu} F^{\mu \nu}\right]\psi_i,
$$
with
$
iD^\mu=i \partial^\mu-g_v V^\mu-
\frac{g_\rho}{2}\boldsymbol\tau \cdot \mathbf{b}^\mu - e A^\mu
\frac{1+\tau_3}{2},
$
$
M^*_p=M^*_n=M^*=m-g_s\phi,
$
and the mesonic and photonic terms  defined as in \cite{fang17}. The term $\mathcal{L}_{\omega \rho } = \Lambda_v g_v^2 g_\rho^2 V_{\mu }V^{\mu }
\mathbf{b}_{\mu }\cdot \mathbf{b}^{\mu }$ couples the $\rho$ to the
$\omega$ meson and allows the  softening of  the density dependence of the
symmetry energy above saturation density  \cite{hor01,pais16V}.
We consider the NL3 \cite{nl3}, and NL3$\omega\rho$
\cite{hor01,pais16V}  parametrizations, which describe  two solar
mass stars \cite{fortin16}.
The last
 ones are obtained from the NL3 model by
 including the $\omega\rho$ term. All models have the same isoscalar properties at saturation, in
particular, the binding energy  $E_b=-16.2$ MeV, the saturation
density $\rho_0=0.148$ fm$^{-3}$, and the incompressibility $K= 272$ MeV. 
The isovector properties, such as  the  symmetry energy and its
slope $L$ at saturation, vary from model to model, and have been fixed such that, at $\rho=0.1$ fm$^{-3}$, all models have the same symmetry energy, $\epsilon_{sym}(0.1)=25.7$ MeV. Besides NL3 with
$L=118$ MeV, we also take NL3$\omega\rho$ with $L=88, \,68$ and 55
MeV. The model with $L=55$ MeV satisfies the constraints imposed by
microscopic calculations of neutron matter \cite{neutron}.
The nucleon AMM is introduced via the coupling of the baryons to the
electromagnetic field tensor with $\sigma_{\mu\nu}=\frac{i}{2}\left[\gamma_{\mu}, \gamma_{\nu}\right] $,
and strength $\kappa_{b}$, with $\kappa_{n}=-1.91315$ for the neutron,
and $\kappa_{p}=1.79285$ for the proton. $\mu_N$ is the nuclear
magneton. We will not consider the AMM of the electrons because its
contribution is negligible  for the magnetic field intensities we
consider in the present work \cite{duncan00}.

The state which minimizes the energy of asymmetric $npe$ matter is characterized by the distribution functions
$
f_{0 i \pm}= [1+e^{(\epsilon_{0i} \mp \nu_i)/T}]^{-1},
$
with 
$\nu_i=\mu_i - g_v V_0  - \frac{g_\rho}{2}\, \tau_i b_0$ for  $i=p,n$, and  $\nu_e=\mu_e$ for the electrons,
and by the  constant mesonic fields which obey the mesonic equations \cite{fang17}.
For $T=0$ MeV, the distribution functions $f_{0i\pm}$ become
$f_{0i+}=\theta(P_{Fi}^2-p^2)$, $f_{0i-}= 0$  \cite{brito06}.

Nuclear matter at subsaturation densities has a liquid-gas phase
transition.  Homogeneous matter is unstable  if the free energy
curvature is negative. The stability conditions for asymmetric nuclear
matter are obtained from the free energy density,  by imposing  
that the function is convex on the densities $\rho_p$ and
$\rho_n$, keeping the volume and temperature constant
\cite{chomaz03}. 
 The thermodynamical spinodal is the surface in the
 ($\rho_n$, $\rho_p$, $T$) space  where the
 determinant of the  free energy curvature matrix 
is zero. Inside this surface, nuclear matter is unstable.

{\it Symmetry energy effect:} We first discuss the effect of the
symmetry energy on the crust-core transition. 
The density and the proton fraction of the crust-core
transition in a neutron star {are functions of} the density dependence of the
symmetry energy. In particular, they  are  correlated with the slope $L$ of
the symmetry energy at saturation
\cite{vidana09,xu09,ducoin10,ducoin11,newton13}. We may, therefore,
expect that the effect of a strong magnetic field on the transition will also depend on the
symmetry energy, since the magnetic field is sensitive to the amount
of protons: the smaller the proton density, the stronger are the
effects.  In previous studies \cite{fang16,fang17}, this aspect
has already been identified. 
\begin{figure}[!t]
\includegraphics[width=0.85\linewidth]{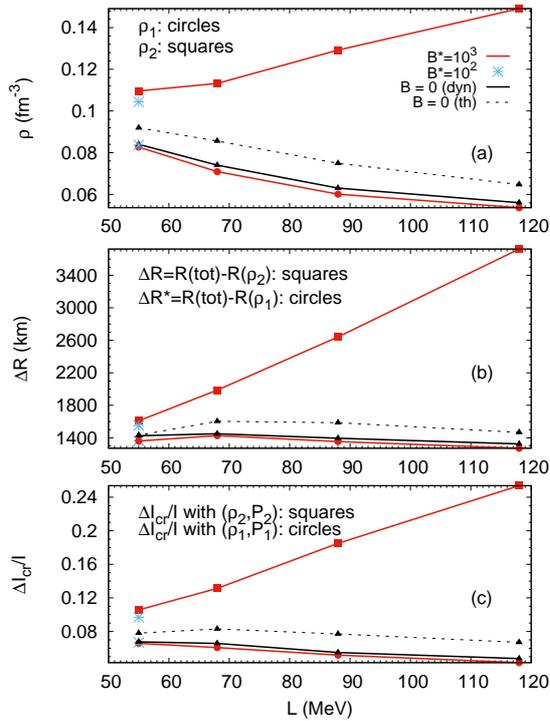} \\
\caption{Transition densities, $\rho_1$ and $\rho_2$ (a); the crust thickness, $\Delta R$ 
  and $\Delta R^*=R({\rm tot})-R(\rho_1)$ (b); the crust fractional momentum of inertia, calculated
with $(\rho_1,P_1)$ and with $(\rho_2,P_2)$ (c)
versus the symmetry energy slope $L$, obtained at $T=0$ with $B^*=10^3$
(red) and $B=0$ (black solid), within the dynamical
 spinodal formalism including the AMM.   For $L=55$ MeV  also $B^*=10^2$ is shown
 (blue stars).
$B=0$ results from the thermodynamical spinodal calculation
are also included  (black dashed). }
\label{L}
\end{figure}

 Within the dynamical spinodal formalism presented in \cite{fang17},
 we  determine the maximum growth rates $\Gamma$ as a function of
 the density, using the $B=0$ proton fraction ($y_p^0$) below the crust core transition and,
 above it, the  $\beta$-equilibrium proton fraction.
 Thomas-Fermi (TF) calculations of the inner crust indicate
 that from $\rho\sim 0.01$ fm$^{-3}$  up to the crust-core transition
 density, which at $B=0$ we designate by $\rho_t^0$, the
 proton fraction does not change much \cite{grill12}.
Unlike the case for $B=0$, there is no well defined transition density
for a strong magnetic field, but a sequence of unstable and stable regions
ranging from $\rho_1$ -- defined, as in \cite{fang17}, as the first time
the growth rate falls to zero, which is smaller than but close to $\rho_t^0$ -- up
to $\rho_2$ -- the onset of the homogenous matter, taking the proton fraction
of $\beta$-equilibrium matter. Both densities coincide with $\rho_t^0$ at $B=0$.


The 
four models introduced above have the same isoscalar properties, but a different
density dependence of the symmetry energy. In Fig. \ref{L}, we show,
as a function of the slope $L$: a) the  densities that define the beginning
  and the end of the transition region, $\rho_1$ and $\rho_2$  (a); b) the  thickness of the crust calculated with $\rho_2$,
  $\Delta R=R(0)-R(\rho_2)$, and with $\rho_1$,
$\Delta R^*=R(0)-R(\rho_1)$ (b);
 and c) the crust fractional moment of inertia,  $\Delta I_{cr}/ I$,  using the approximate expression \cite{Lattimer00}
\begin{eqnarray}
\frac{\Delta I_{cr}}{I}&\simeq& \frac{28\pi P_t
                                R^3}{3M}\frac{(1-1.67\beta-0.6\beta^2)}{\beta}
                                \nonumber \\
                       &\times&\left[1+\frac{2P_t(1+5\beta-14\beta^2)}{\rho_t
                                m\beta^2}\right]^{-1} \, ,
\label{Icr}
\end{eqnarray}
and taking for the transition density $\rho_t$ and pressure $P_t$ the limiting densities of
the transition density, ($\rho_2$, $P_2$) and ($\rho_1$, $P_1$) (c).
In this expression, $\Delta I_{cr}$ is the crust moment of inertia, $I$ is the
total moment of inertia of the star, $M$ and $R$ are the gravitational mass and
radius of the star, $\beta=GM/R$ is the compactness parameter, and
$m$ is the nucleon mass.  
The quantities in Fig. \ref{L} are  calculated at $T=0$  for
$B^*=10^3$ and $L=55$ 
from  the dynamical spinodal formalism with AMM. The blue stars are
for  $B^*=10^2$. For $B=0$, we  include the results from a dynamical
and a thermodynamical spinodal calculation, respectively, with and
without AMM.

The effect of $B$ and $L$
on the thickness of the crust is summarized in the following: a) the larger the $L$, the larger the
effect of $B$, mainly due to the proton fraction associated
with each model, since  a larger $L$ is associated with a smaller
proton fraction;
b) compared to $B=0$, the effect can
be as large as a 100\% 
for $L=118$~MeV. However, experimental contraints \cite{tsang12} and microscopic
neutron matter calculations \cite{neutron} indicate that the models with $L=30-80$
MeV are more realistic. For $L=55$ MeV, the effect 
corresponds to an increase of
$\sim 20\%$; c) the lower limit of the crust-core transition defined
by $\rho_1$ is just slightly smaller than the $B=0$ crust-core
transition $\rho_t^0$. The magnetic field  essentially creates a
complex transition region above this density;
d) taking $L=55$ MeV and decreasing the magnetic field by
an order of magnitude from $B^*=10^3$ to $B^*=10^2$, quantities such as
the transition density, the crust thickness and the crust fraction of
moment of inertia, defined with the density $\rho_2$, suffer a reduction of $\sim  3-5\%$, but are still
 larger than the corresponding quantities at $B=0$.
We conclude by stressing that properties of magnetized neutron stars that directly depend on the thickness
of the crust may set stringent constaints on the symmetry energy slope $L$.

\begin{figure}[!th]
\includegraphics[width=0.85\linewidth]{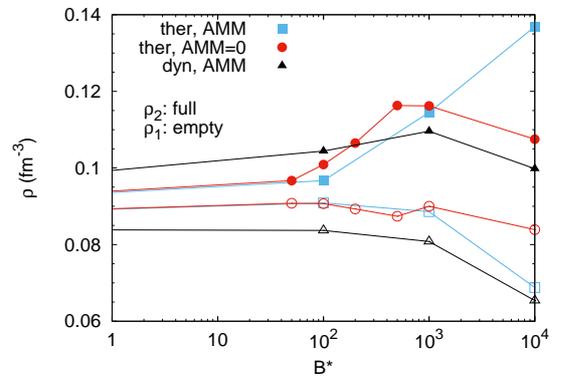} \\
\caption{The transition densities, $\rho_1$ (empty) and $\rho_2$ (full) obtained with the $L=55$ MeV model, at $T=0$, for several values of $B^*$, and using the  thermodynamical spinodal formalism with (squares) and without AMM (circles), and the dynamical
 spinodal with AMM (triangles).  } 
\label{fig1a}
\end{figure}

\begin{figure}[!t]
\includegraphics[width=0.9\linewidth]{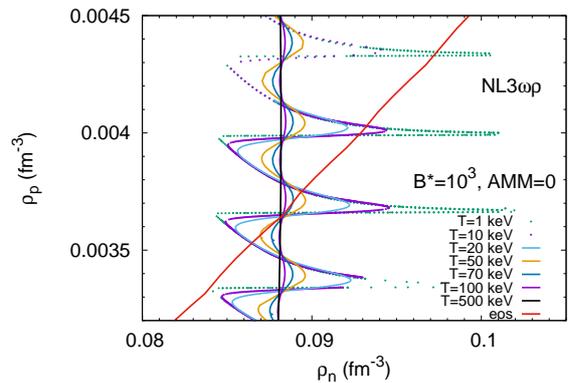} \\
\caption{Details of the crossing of the thermodynamical spinodal with the EoS (black solid line) for NL3$\omega\rho$ with $B^*=10^3$, considering different temperatures, and taking AMM$=0$. }
\label{cross}
\end{figure}

{\it Temperature effect:} We estimate the effect of
temperature on the crust transition by calculating the thermodynamical
spinodal of strongly magnetized  nuclear matter.  In Ref. \cite{avancini06}, it has been shown that due to the
large incompressibility of the electron gas, most models that describe
$npe$ matter  do not present thermodynamical instabilities, or present
only a very reduced region of instabilities. 
Thermodynamical stability does not necessary mean that the $npe$ system is
stable to small density fluctuations, as shown in
\cite{pethick95,providencia06,brito06}. Calculating the  dynamical
spinodal  determines precisely the instability region taking into
account the independent fluctuations of the neutron, proton and
electron densities.
 However, according to  \cite{avancini10,ducoin11},   the $np$
matter thermodynamical spinodal  gives a good prediction of  the
crust-core transition density, just slightly above the prediction from
a TF calculation or a dynamical spinodal for $npe$ matter. 
This behavior is confirmed in Fig. \ref{L} 
where 
the $B=0$
quantities determined from the dynamical and the thermodynamical
spinodals have been plotted. The values predicted from the thermodynamical
spinodal are always $\sim
15\%$ larger than the ones from the dynamical spinodal. For a strong
magnetic field with an intensity of the order we have considered in this work, the effect is
similar. 
In Fig. \ref{fig1a},  we plot the
crust-core transition  densities, $\rho_1$ and $\rho_2$, obtained at
$T=0$ with the $L=55$ MeV model from the $npe$ dynamical  spinodal with AMM, 
and from  the $np$ thermodynamical spinodals with and without AMM, to
estimate the limitations of our predictions.  The lower (upper) density $\rho_1$ ($\rho_2$) corresponds to the density
where the  $\beta$-equilibrium  EoS first (last) crosses the
spinodal, see  Fig. \ref{cross}.

Comparing the results obtained from the dynamical
and thermodynamical spinodals we conclude the following:
a) the dynamical
and thermodynamical spinodals predict  the same trends for the
transition densities, though the dynamical spinodal predicts smaller
values of $\rho_1$, in accordance with results from \cite{avancini10,ducoin11}.
 However, for the upper limit of the transition region, there is a
 dependence on $B$, and the dynamical $\rho_2$ is larger (smaller) than the
 thermodynamical one for $B^*<10^2$ ($B^*>10^2$);
b)  AMM does not affect much the results obtained with $B^*<10^3$.  However,  the AMM reduces in a non-negligible way the
instability region for the larger fields, giving rise to smaller crust
thicknesses and momentum of inertia crustal fractions. 

The temperature of the crust decreases as the star cools. While a very
young star, less than one year old, may have a inner crust temperature above
$10^9$ K, it will drop below 
$10^9$ K , or even $10^8$ K, depending on the EoS considered and the
mass of the star \cite{chamel08,yakovlev01}. It is, therefore,
reasonable to ask whether the strong effect of the magnetic field on
the crust-core transition calculated at $T=0$,  with the appearance of a  transition region
where stable and unstable regions alternate, still persists at finite
temperature.
Moreover, the time
evolution of both the magnetic field and temperature  inside the star
are strongly coupled, and, therefore, it is important to understand
which is the effect of the temperature on the transition region created by
a magnetic field.

We  calculate the crust-core  transition
density/region for temperatures in the range  1
keV$< T<  $ 1 MeV ($10^7\lesssim T \lesssim 10^{10}$ K)
from the
thermodynamical spinodal without AMM. Above
$B^*\sim 10^3$ ($B\sim 5\times 10^{16}$ G), the AMM has a non-negligible effect and, therefore, we will
essentially restrict ourselves to values below that number.
{ As discussed in \cite{fang16,fang17}, 
  the spinodal section shows a complex structure and bands of
  instability with large isospin asymmetry appear 
associated with the filling of the different Landau levels.  As a result}
for low temperatures, the $\beta$-equilibrium EoS crosses the
spinodal section several times, defining the region of instability
referred before, see  Fig. \ref{cross}. 
The transition region for $T=10$ keV is smaller than for $T=1$ keV since the EoS is not crossing  the
last band shown. The transition region decreases as $T$ increases and,
for large enough temperatures, the crossing occurs at a well defined
density, as for $T\ge 100$ keV in  Fig. \ref{cross}.
{ The Landau quantization will be completely washed out by
  temperatures 
of the order of the energy separation between consecutive Landau
levels, i.e.
$T\gtrsim {eB}/{M^*}= {m_e^2B^*}/{M^*}$. For $B^*=1000$ and
taking $M^*\sim700$MeV for $\rho\sim 0.09$~fm$^{-3}$, this corresponds
to $T\gtrsim 0.3$~MeV. The effects become important already for 10\% of
 this value in the regions of larger isospin asymmetry, e.g. larger $\rho_n$.}

\begin{figure}[!th]
\includegraphics[width=0.75\linewidth]{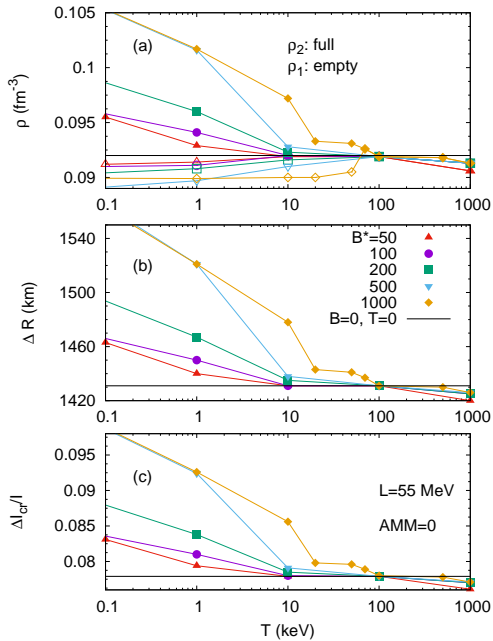} \\
\caption{The transition densities, $\rho_1$ (empty) and $\rho_2$ (full), (a), the crust thickness  (b), and
the momentum of inertial  crustal fraction (c) for NL3$\omega\rho$, with
 $L=55$ MeV,  for several values of $B^*$ and $T$.} 
 \label{fig3}
\end{figure}

We plot in Fig. \ref{fig3} the
transition densities, $\rho_1$ and $\rho_2$ (a), the crust
thickness, $\Delta R$ (b), and the momentum of inertia
crustal fraction (c) for $B^*\le 10^3$ and $10^{-1}\le T\le
10^3$ keV . {These quantities together with the corresponding
  transition pressures are given in the supplementary material \cite{supmat}.}
The crust thicknesses are estimated from the Tolmann-Oppenheimer-Volkoff (TOV)
equations  \cite{tov} at $B=0$. 
The densities
$\rho_2$ come closer to the lower limit, $\rho_1$, of the transition region as the temperature increases, and for the magnetic field intensities considered, all magnetic field
effects have been washed out at $T=100$ keV,
and the $B=0$ transition density  has been recovered. For a stronger field, this is not anymore
true, but since for these stronger fields, several of the suppositions
considered in the present work break, such as the use of the TOV
equations or the exclusion of the AMM of the nucleons, we will not
discuss so strong fields. Above $T=100$ keV, $\rho_1$ and $\rho_2$
coincide, but they take values below the $T=0$ transition density: this is the
reduction of the extension of the spinodal section due to
 temperature effects. To summarize, we may expect the appearance of a
 transition region {with nonzero thickness for crustal temperatures} below 100 keV and a magnetic field intensity at the crust-core
 transition below $B\sim 5\times 10^{16}$ G.

 Also, the crust momentum of inertia fraction is affected, and is
 large enough to account for the Vela glitches, which,  and according to
 \cite{glitch2}, would require a fractional crustal momentum of
 inertia of the order of $\sim 0.065-0.095$, considering that the effective neutron
 mass, including entrainment effects, is $4-6$ times larger than the
 neutron bare mass. However, further studies should be undertaken
 because strong magnetic fields as the ones considered in the present
 work will certainly influence the neutron superfluid behavior and
 affect the neutron entrainment to the lattice.

The main effect of having used the thermodynamical spinodal instead of the
dynamical one is that the predicted crust-core transition density  is $\sim
10\%$ larger, the crust fraction momentum of inertia $\sim 10-15\%$
larger, and the transition region slightly smaller, but the overall conclusions remain valid.

{\it Conclusion:} We have analyzed how the effects of a strong magnetic field on the neutron star crust, previously studied in \cite{fang16,fang17}, are affected by the density dependence of the symmetry energy of the EoS,  and by the temperature of the crust, within a RMF description of $npe$ and $np$ matter. At $T=0$,  the crust-core transition was obtained from the dynamical spinodal with AMM, and at finite temperatures,  from the thermodynamical spinodal, excluding the AMM, which is justified since only magnetic fields below $5\times 10^{16}$ G are considered.

We have confirmed the results of Refs.  \cite{fang16,fang17}: 
due to the sensitivity of the magnetic field to the proton density, the extension of the  crust-core transition region strongly depends on the slope $L$ of the symmetry energy. The larger the slope $L$, the larger the transition region, because, below saturation density, models with a large $L$ present smaller symmetry energies and, therefore, accept smaller proton fractions. Experimental and theoretical constraints seem to limit $L$  below 80 MeV (30$<L<80$MeV) \cite{tsang12}, resulting in a more moderate  effect of the magnetic field on the extension of the crust. Properties of magnetized neutron stars that directly depend on the thickness
of the crust can set stringent constraints on the symmetry energy slope $L$ due to the great sensitivity of the crust size to this property.

We have also studied the effect of temperature for  magnetic fields $B\le 5\times 10^{16}$ G. The magnetic field effects on the extension of the  transition density are washed out for temperatures above $10^9$ K, but below these temperatures, even a field of intensity $2\times 10^{15}$ G will have a finite effect on the crust thickness. Microphysical parameters, such as transport coefficients, that enter in the magneto-thermal evolution equations of a neutron star, are certainly affected by the existence of the crust-core transition region that changes with cooling, and the impact of this effect should be investigated. Recently, a  one dimensional thermal-magneto-plastic model, that considered transport coefficients sensitive to temperature, as well as the coupling of the crustal motion to the magnetosphere, has been implemented, and it has been shown that this coupling induces an enrichment  and acceleration of the magnetar  dynamics \cite{li2016}.

\bigskip
{\it Acknowledgments}:
This work  is  partly  supported  by
the FCT (Portugal) project UID/FIS/04564/2016,
and by ``NewCompStar'', COST Action MP1304. H.P. is supported by FCT
(Portugal) under Project No. SFRH/BPD/95566/2013.


\end{document}